\documentclass[useAMS,usenatbib,usegraphicx]{mn2e}
\usepackage{color}
\usepackage{wasysym}
\usepackage{url}
\usepackage{multirow}

\def\kms{km\,s$^{-1}$}

\def\SiII{Si\,{\sc ii}}
\def\SiIII{Si\,{\sc iii}}

\def\MgII{Mg\,{\sc ii}}
\def\CaII{Ca\,{\sc ii}}

\def\FeII{Fe\,{\sc ii}}

\def\CoII{Co\,{\sc ii}}

\def\CaII{Ca\,{\sc ii}}

\def\dm15{$\Delta m_{15}(B)$}

\def\lesssim{\mathrel{\hbox{\rlap{\hbox{\lower4pt\hbox{$\sim$}}}\hbox{$<$}}}}

\def\gtrsim{\mathrel{\hbox{\rlap{\hbox{\lower4pt\hbox{$\sim$}}}\hbox{$>$}}}}

\def\aj{AJ}%
\def\apj{ApJ}%
\def\apjl{ApJ}%
\def\apjs{ApJS}%
\def\aap{A\&A}%
\def\mnras{MNRAS}%
\def\pasp{PASP}%
\title[Near-infrared spectra of SN~2005cf]
{
  Interpreting the
  near-infrared spectra of the 'golden standard' Type Ia supernova
  2005cf\thanks{Based on observations collected at ESO,
    Paranal. Program ID 075.D-0823(B). 
    }
}

\author[E.~E.~E.~Gall~et~al.]
{E.~E.~E.~Gall,$^{1}$\thanks{E-mail: egall@mpa-garching.mpg.de}
 S.~Taubenberger,$^{1}$
 M.~Kromer,$^{1}$
 S.~A.~Sim,$^{2}$
 S.~Benetti,$^{3}$
 \newauthor
 G.~Blanc,$^{4,3}$
 N.~Elias-Rosa$^{5,3}$ and
 W.~Hillebrandt$^{1}$\\
 $^{1}$Max-Planck-Institut f{\"u}r Astrophysik, 
       Karl-Schwarzschild-Str. 1, 85741 Garching bei M{\"u}nchen, Germany\\
 $^{2}$Research School of Astronomy \& Astrophysics, 
       Mount Stromlo Observatory, Cotter Road, Weston ACT 2611, Australia\\
 $^{3}$INAF Osservatorio Astronomico di Padova, 
       Vicolo dell'Osservatorio 5, 35122 Padova, Italy\\
 $^{4}$Universit\'e Paris Diderot-Paris 7,
       Laboratoire APC, 10 rue Alice Domon et L\'eonie Duquet, 75205 Paris cedex 13, France\\
 $^{5}$Institut de Cincies de l'Espai (IEEC-CSIC), 
       Facultat de Cincies, Campus UAB, 08193 Bellaterra, Spain
}

\begin{document}

\date{\today}

\pagerange{\pageref{firstpage}--\pageref{lastpage}} \pubyear{2012}

\maketitle

\label{firstpage}

\begin{abstract}
  We present nine near-infrared (NIR) spectra of supernova (SN) 
  2005cf at epochs from $-10$\,d to $+42$\,d with respect to 
  $B$-band maximum, complementing the existing excellent data sets 
  available for this prototypical Type Ia SN at other wavelengths. 
  The spectra show a time evolution and spectral features 
  characteristic of normal Type Ia SNe, as illustrated by a 
  comparison with SNe~1999ee, 2002bo and 2003du. The broad-band 
  spectral energy distribution (SED) of SN~2005cf is studied in 
  combined ultraviolet (UV), optical and NIR spectra at five epochs 
  between $\sim$\,8\,d before and $\sim$\,10\,d after maximum light. 
  We also present synthetic spectra of the hydrodynamic explosion 
  model W7, which reproduce the key properties of SN~2005cf not 
  only at UV-optical as previously reported, but also at NIR 
  wavelengths. From the radiative-transfer calculations we infer 
  that fluorescence is the driving mechanism that shapes the 
  SED of SNe~Ia. In particular, the NIR part of the spectrum is 
  almost devoid of absorption features, and instead dominated by 
  fluorescent emission of both iron-group material and 
  intermediate-mass elements at pre-maximum epochs, and pure 
  iron-group material after maximum light. A single P-Cygni feature 
  of \MgII\ at early epochs and a series of relatively unblended 
  \CoII\ lines at late phases allow us to constrain the regions of 
  the ejecta in which the respective elements are abundant.

\end{abstract}

\begin{keywords}
supernovae: individual: SN~2005cf -- supernovae: individual: SN~1999ee -- supernovae: individual: SN~2002bo -- supernovae: individual: SN~2003du -- techniques: spectroscopic -- 
radiative transfer
\end{keywords}

\section{Introduction}

An empirical relation between their maximum brightness and light curve
decline rate \citep{phillips1993a} makes Type Ia supernovae (SNe~Ia)
one of the best tools to measure cosmological distances. Using this
approach \citet{riess1998a} and \citet{perlmutter1999a} discovered
that our Universe undergoes an accelerated expansion. This was in turn
interpreted as evidence for the existence of `dark energy'
(e.g.~\citealt{riess2007a}). Despite this importance for astrophysics
and cosmology our knowledge about the progenitor systems or the
explosion mechanisms of these luminous events is still limited. They 
are commonly
believed to be thermonuclear explosions of carbon-oxygen white dwarfs
in binary systems. However, many aspects are still unclear; e.g. the
masses of the exploding stars, the nature of the companion star, the
ignition conditions, the exact mechanism of the burning process
or the origin of the various subclasses of SNe~Ia. Systematic 
uncertainties arising from this ignorance are one factor limiting 
the precision of cosmological measurements. One way to improve our
understanding of these objects is the comparison of observed spectra
with predictions from theoretical models. This, however, requires
complete data sets with a wide wavelength coverage \citep{roepke2012a}.

In particular, the near-infrared\footnote{In this paper we will refer 
  to wavelengths between 9500 and 26\,000\,\AA\ when talking about
  `near-infrared'.} (NIR) regime has been proposed as a useful tool 
to discriminate between different explosion models 
\citep[e.g.][]{marion2003a}. This is mainly due to the fact that the 
lower opacity in the NIR enables us to look deeper into the ejecta 
and thus probe layers still hidden to optical observations at the 
same epochs. Also, the NIR is believed to be less severely affected 
by line blending compared to optical or ultraviolet (UV) wavelengths 
\citep{harkness1991,hoeflich1993a,spyromilio1994,wheeler1998}. 
At the same time, however, observations suggest SNe~Ia to be 
excellent NIR standard candles (\citealt{elias1985a,meikle2000,
krisciunas2004a}, but see \citealt{kattner2012a}), in contrast to 
optical wavelengths where they can only be used as 
standard\textit{izeable} candles. This result was confirmed by a series 
of toy explosion models \citep{kasen2006b}. While the homogeneity in 
the NIR might point towards a common explosion mechanism for all 
SNe~Ia, it could also mean that the NIR photons are not as sensitive 
to variations in the ejecta as previously suggested.

While optical spectra have been studied quite extensively, the NIR
regime is still comparatively unexplored. Although there are quite 
a few publications containing NIR spectra of SNe~Ia (e.g. \citealt{frogel1987a,spyromilio1992a,spyromilio2004a,meikle1996a,
bowers1997,hamuy2002a,hamuy2002b,hoeflich2002b,rudy2002,marion2003a,
marion2009,benetti2004a,eliasrosa2006a,stanishev2007b,taubenberger2011})
one has to realize that for many SNe there is only one spectrum
available -- typically at post-maximum epochs -- or that the objects
are of a peculiar subclass. The number of well-sampled \textit{normal}
SNe~Ia with pre-maximum-light observations in the NIR is actually
quite small. \citet{meikle1996a} have published six spectra of SN~1994D,
\citet{hamuy2002a,hamuy2002b} eleven spectra of SN~1999ee, 
\citet{benetti2004a} six spectra of SN~2002bo, 
\citet{pignata2008} four spectra of SN~2002dj, 
\citet{eliasrosa2006a} ten spectra of SN~2003cg 
and \citet{marion2009} four spectra of SN~2005am.
Moreover, there are 14 spectra of SN~2003du out to the nebular phase,
published by \citet{motohara2006}, \citet{stanishev2007b} and 
\citet{marion2009}.

In order to increase the sample of SNe~Ia with good NIR coverage we
present and analyze a set of nine NIR spectra of SN~2005cf, ranging
from $-10.2$\,d to $+41.5$\,d with respect to $B$-band
maximum. SN~2005cf, according to \citet{wang2009} `the golden standard
Type Ia supernova', is a spectroscopically normal SN~Ia with a more
complete data set than most other SNe~Ia. Optical and NIR photometry
as well as optical spectroscopy of SN~2005cf have been published by
\citet{garavini2007}, \citet{pastorello2007a} and \citet{wang2009}. UV
photometry and spectroscopy have been published by \citet{bufano2009}
and \citet{wang2012a}.

Table \ref{table:SN2005cf_properties} summarizes the most important
properties of SN~2005cf: at $B$-band maximum, it had a magnitude of
$M_{B_{\mathrm{max}}} = -19.39$ \citep{pastorello2007a}, $\Delta
\mathrm{m}_{15}(B)$ lies between 1.07 \citep{wang2009} and 1.12
\citep{pastorello2007a}, and the derived \textsuperscript{56}Ni mass
is about 0.7\,$\mathrm{M}_{\astrosun}$ \citep{pastorello2007a} or
0.78\,$\mathrm{M}_{\astrosun}$ \citep{wang2009}. This corresponds
quite well to the average values obtained for normal SNe~Ia. An important
shortcoming of the data set of SN~2005cf has so far been the lack of
published NIR spectroscopy. This deficiency is now remedied with the
data presented in this work.

\begin{table}
  \caption{Properties of SN~2005cf, as derived in the literature.}
  \centering
  \begin{tabular}{@{}lcc@{}}
  \hline
  & \citeauthor{pastorello2007a} & \citeauthor{wang2009} \\
  & \citeyearpar{pastorello2007a} & \citeyearpar{wang2009} \\
  \hline
  JD of $B_{\mathrm{max}}$ & 2\,453\,534.0 $\pm$ 0.3& 2\,453\,533.66 $\pm$ 0.28 \\
  $m_{B_{\mathrm{max}}}$ & 13.54 $\pm$ 0.02 & 13.63 $\pm$ 0.02 \\
  $M_{B_{\mathrm{max}}}$ & $-19.39$ $\pm$ 0.33 &  \\
  $\Delta \mathrm{m}_{15}(B)_{\mathrm{true}}$ & 1.12 $\pm$ 0.03 & 1.07 $\pm$ 0.03 \\
  $M(^{56}\mathrm{Ni})$ & 0.7\,$\mathrm{M}_{\astrosun}$ & (0.78 $\pm$ 0.10)\,$\mathrm{M}_{\astrosun}$ \\
  $\mu$(MCG-01-39-003) & (32.51 $\pm$ 0.33)\,mag &  (32.31 $\pm$ 0.11)\,mag \\
  $E(B - V)_{\mathrm{host}}$ & 0\,mag & (0.10 $\pm$ 0.03)\,mag \\
  \hline
  \end{tabular}
  \label{table:SN2005cf_properties}

\end{table}

The paper is organised as follows. In Section~\ref{Observations} the
NIR spectra of SN~2005cf are presented and the process of data
reduction is described. In Section~\ref{Discussion} we compare these
observations to the NIR spectra of other SNe~Ia and construct combined 
UV-optical-NIR spectra to investigate the broad-band spectral energy 
distribution (SED). Synthetic spectra of the hydrodynamic explosion 
model W7 \citep{nomoto1984a} are used to explain the SED in terms of 
fluorescence and to perform a NIR line identification 
(Section~\ref{Comparison_SN2005cf_W7}). Conclusions are drawn in
Section~\ref{Conclusions}.

\section{Observations and data reduction}
\label{Observations}

The spectroscopic data of SN~2005cf were obtained with ISAAC \citep{moorwood1999}, mounted at the Very Large Telescope at
the Paranal Observatory of ESO (European Southern Observatory), and
NICS (Near-Infrared Camera Spectrometer), mounted at the Telescopio
Nazionale Galileo of Fundaci\'{o}n Galileo Galilei and INAF (Istituto
Nazionale di Astrofisica). The spectra were collected by the European
Research Training Network (RTN) `The Physics of Type Ia Supernova
Explosions'\footnote{http:/$\!$/www.mpa-garching.mpg.de/\url{~}rtn/}. They
cover the time interval from 10\,d before to 42\,d after $B$-band
maximum (see Table \ref{table:SN2005cf_epochs} for details).

\begin{table}
  \caption{Epochs of the NIR spectra of SN~2005cf.}
  \centering
  \begin{tabular}{lrrl}
  \hline
  UT date & Julian Day & Epoch & Instrument \\
  \hline
  2005/06/02 & 2\,453\,523.8 & $-10.2$\,d & ISAAC \\
  2005/06/03 & 2\,453\,524.5 &  $-9.5$\,d & NICS \\
  2005/06/07 & 2\,453\,528.6 &  $-5.4$\,d & ISAAC \\
  2005/06/12 & 2\,453\,533.7 &  $-0.3$\,d & ISAAC \\
  2005/06/14 & 2\,453\,535.8 &  $+1.8$\,d & ISAAC \\
  2005/06/23 & 2\,453\,544.7 & $+10.7$\,d & ISAAC \\
  2005/06/26 & 2\,453\,548.4 & $+14.4$\,d & NICS \\
  2005/07/13 & 2\,453\,565.4 & $+31.4$\,d & NICS \\
  2005/07/24 & 2\,453\,575.5 & $+41.5$\,d & NICS \\
  \hline
  \end{tabular}
  \\[1.5ex]
  \flushleft
  ISAAC = 8.2\,m Very Large Telescope + ISAAC, short-wavelength 
  low-resolution spectroscopy (SWS1-LR); \, 
  http:/$\!$/www.eso.org/sci/facilities/paranal/instruments/isaac/\\
  \vspace{0.13cm}
  NICS = 3.58\,m Telescopio Nazionale Galileo + NICS, Amici prism; 
  \,http:/$\!$/www.tng.iac.es/instruments/nics/\\
  \label{table:SN2005cf_epochs}
\end{table}

The spectra taken with ISAAC (at epochs $-10.2$\,d, $-5.4$\,d,
$-0.3$\,d, 1.8\,d and 10.7\,d with respect to $B$-band maximum) were
taken in the short-wavelength low-resolution setup (resolution 
$R = \frac{\lambda}{\Delta\lambda} \approx 500$). The
observational data are divided into the four bands $SZ$, $J$, $SH$ and
$SK$, which cover the wavelength range from 9000\,\AA\ to
25\,600\,\AA.  The NICS spectra (at epochs $-9.5$\,d, 14.4\,d, 31.4\,d
and 41.5\,d with respect to $B$-band maximum) were taken with an Amici
prism as disperser. They cover the whole NIR region between about
7500\,\AA\ and 26\,000\,\AA\ at very low resolution ($R \approx 50$).

For the NICS data, cross-talk and distortion corrections were applied
with \textsc{snap}\footnote{Speedy Near-IR data Automatic reduction
  Pipeline,\\ http:/$\!$/www.arcetri.astro.it/\url{~}filippo/snap/}.
All other reduction steps for both the NICS and the ISAAC data were
performed using \textsc{iraf}\footnote{\textsc{iraf} (Image Reduction
  and Analysis Facility) is distributed by the National Optical
  Astronomy Observatories, which are operated by the Association of
  Universities for Research in Astronomy, Inc., under cooperative
  agreement with the National Science Foundation.}, following standard
procedures in the NIR. This included dark-current and flat-field
corrections, and a pairwise subtraction of sub-exposures taken with
the target off-set along the slit in order to remove the sky emission.
One-dimensional spectra were obtained from the two-dimensional images
using an optimal, variance-weighted extraction method
\citep{horne1986a}.  For the ISAAC spectra the wavelength was
calibrated with the help of arc-lamp spectra taken along with the SN
exposures using exactly the same telescope configuration, and the
calibration was checked with the position of known telluric features.
For the NICS spectra, the wavelength calibration was accomplished with
tabulated values. Telluric features were removed from the SN spectra
by dividing through spectra of telluric standard stars of type A0 or
similar, which had been observed at a similar airmass as the SN. A
first relative flux calibration was provided by multiplication with a
Vega spectrum.  Then the absolute fluxes were adjusted to agree with
contemporaneous $JHK$ photometry by \citet{pastorello2007a} and
\citet{wang2009}.  Finally, redshift ($z=0.00646$) and reddening
[$E(B-V)_\mathrm{Gal}=0.097$ mag, \citep*{schlegel1998a}] corrections
were applied.

\section{Results}
\label{Discussion}

\begin{figure}
 \includegraphics[width=84mm]{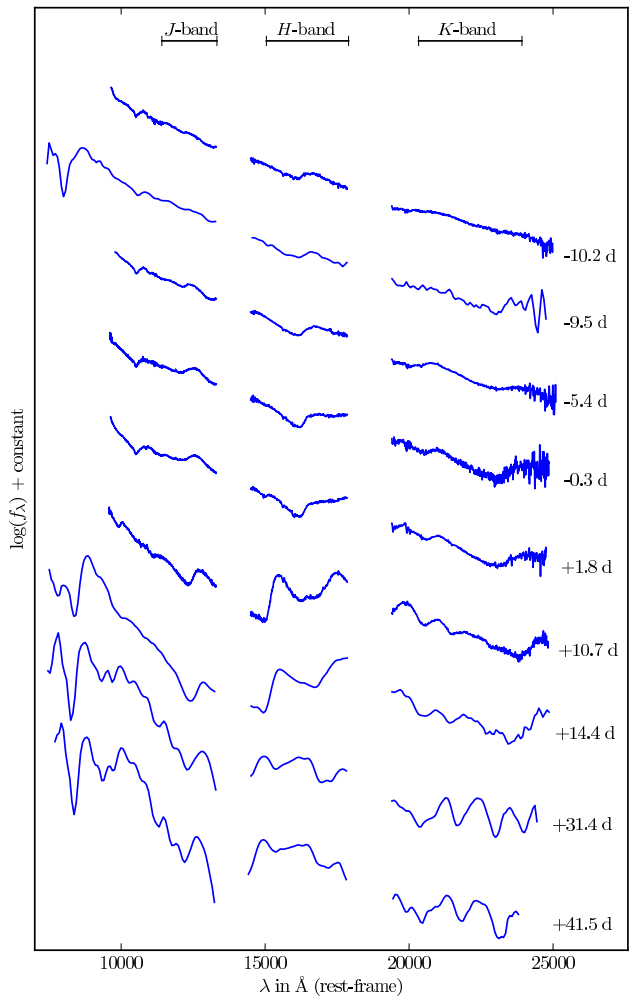}
 \caption{Time sequence of our NIR spectra of SN~2005cf. The gaps 
  between the NIR bands are regions where the Earth's atmosphere is 
  opaque.}
 \label{figure:SN2005cf_time_sequence}
\end{figure}

The nine final, reduced NIR spectra of SN~2005cf are presented in
Fig.~\ref{figure:SN2005cf_time_sequence}. Before maximum light, large
parts of the spectra resemble the infrared tail of a black-body
continuum. This changes significantly at later epochs, when the
spectra start to show increasingly pronounced spectral
features. Around 10\,d after maximum (and thereafter) two distinct
humps form between 15\,000\,\AA\ and 18\,000\,\AA. These have been
previously identified as \CoII\ emission \citep{wheeler1998}.
Characteristic features in the $K$ band, which form at even later 
epochs, have also been attributed to Co emission \citep{wheeler1998}.
Compared to a corresponding black-body continuum there is an increasing 
flux deficit in the $J$-band region after maximum light.

In the following we will take a more detailed look at the spectral
properties of SN~2005cf, comparing the NIR data of SN~2005cf to those
of other well-observed normal SNe~Ia and studying the SED in combined 
UV-through-NIR spectra.

\subsection{Comparison of SN~2005cf to other Type Ia supernovae}
\label{section:Comparison_SN2005cf_other_SNeIa}

Representative, normal SNe~Ia, for which multi-epoch NIR spectroscopy
is available, are e.g. SNe~1999ee \citep{hamuy2002a,hamuy2002b}, 2002bo
\citep{benetti2004a} and 2003du \citep{stanishev2007b}. An overview,
comparing some of their properties to those of SN~2005cf, is given in
Table \ref{table:SN_Comparison}.

\begin{table*}
  \caption{Comparison of characteristic properties of SNe 1999ee,
    2002bo, 2003du and 2005cf.}
  \label{table:SN_Comparison}
  \centering
  \begin{tabular}{lcccc}
   \hline
   & SN~1999ee & SN~2002bo & SN~2003du & SN~2005cf \\
   \hline

   Branch subtype$^{a}$ & `shallow-silicon' & `broad-line' & `core-normal' & `core-normal' \\

   Benetti subtype$^{b}$ & LVG & HVG & LVG & LVG \smallskip \\

   $\dot v$ (\kms\,d$^{-1}$)$^{b}$ & 42 $\pm$ 5 & 110 $\pm$ 7 & 31 $\pm$ 5 & 35 $\pm$ 5 \\

   $v_{10}$(\SiII) ($10^3$\,\kms)$^{b}$ & 9.90 $\pm$ 0.10 & 11.73 $\pm$ 0.15 & 10.10 $\pm$ 0.10 & 9.94 $\pm$ 0.30 \\ 

   $M_{B_{\mathrm{max}}}$ & $-19.85 \pm 0.28$ & $-19.41 \pm 0.42$ & $-19.34 \pm 0.16$ & $-19.39 \pm 0.33$ \\

   $\Delta \mathrm{m}_{15}(B)$ & 0.94 $\pm$ 0.06 & 1.13 $\pm$ 0.05 & 1.02 $\pm$ 0.05 & 1.12 $\pm$ 0.03 \\

   $E(B - V)$ & 0.32\,mag & (0.43 $\pm$ 0.10)\,mag & (0.01 $\pm$ 0.05)\,mag & 0.097\,mag \\

   Redshift $z$ & 0.0117 & 0.0043 & 0.0064 & 0.00646 \smallskip \\ 

   References & \citet{hamuy2002a,hamuy2002b} & \citet{benetti2004a,benetti2005a} & \citet{stanishev2007b} & \citet{garavini2007} \\

     & \citet{stritzinger2002} & \citet{stehle2005a} & \citet{benetti2005a} & \citet{pastorello2007a} \\

     & \citet{benetti2005a} & \citet{branch2006a} & \citet{branch2009} & \citet{branch2009} \\

     & \citet{branch2006a} & & & \citet{wang2009} \\

  \hline

  \end{tabular}
  \\[1.5ex]
  \flushleft
  $^a$~Sub-classification according to \citet{branch2006a}.\quad \\
  $^b$~Low-velocity-gradient (LVG) and high-velocity-gradient (HVG)
  SNe as defined by \citet{benetti2005a}.
\end{table*}

\begin{figure*}
   \includegraphics[width=170mm]{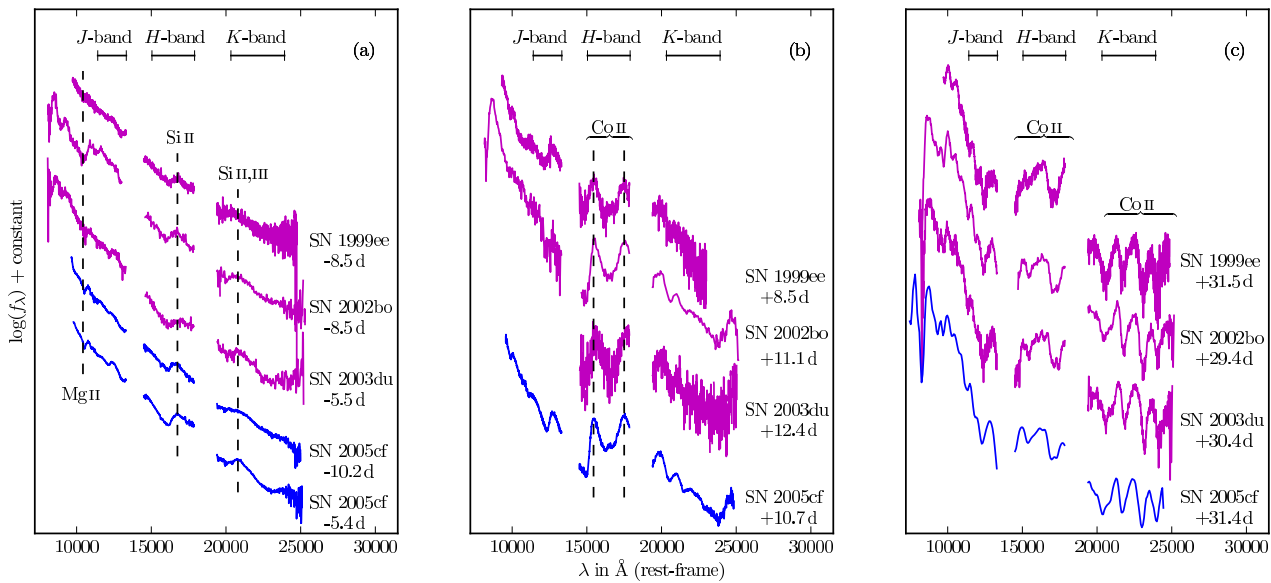}
   \caption{SN~2005cf in comparison with SNe~1999ee, 2002bo and 2003du. 
     Panel (a): about 12 to 5\,d before $B$-band maximum. 
     Panel (b): 9 to 12\,d after $B$-band maximum. 
     Panel (c): 29 to 32\,d after $B$-band maximum.}
  \label{figure:SNe_Ia_NIR_3epochs}
\end{figure*}

The light-curve decline rate $\Delta \mathrm{m}_{15}(B)$ of SN~1999ee
characterizes it as a relatively slowly declining SN~Ia. However, its
peak magnitude agrees with the light-curve shape as described by the
Phillips relation \citep{phillips1993a,phillips1999a}. Marginally
assigned to the `shallow-silicon' class by \citet{branch2006a}, the
spectral features show that, overall, SN~1999ee is a typical SN~Ia
\citep{hamuy2002a,hamuy2002b}. Concerning the spectral features, the
same can be said about SN~2002bo, though at early phases this SN has
higher ejecta velocities than the average normal SN~Ia
\citep{benetti2004a,stehle2005a} and has been included in the
`broad-line' and `high-velocity-gradient' SN~Ia subclasses by 
\citet{branch2006a} and \citet{benetti2005a}, respectively. SN~2003du, 
finally, is very similar to SN~2005cf in many respects, both being 
`core-normal', `low-velocity-gradient' SNe~Ia 
\citep{branch2006a,benetti2005a}. Fig.~\ref{figure:SNe_Ia_NIR_3epochs} 
shows that the NIR spectral evolution of all four SNe is remarkably 
similar, with a few exceptions worth mentioning.

\subsubsection{Spectra ten to five days before $B$-band maximum}
\label{section:SNcomparison_before_Bmax}

Panel (a) of Fig.~\ref{figure:SNe_Ia_NIR_3epochs} shows spectra
between 10.2 and 5.4\,d before $B$-band maximum.

Around 10\,400\,\AA\ there is a very prominent absorption feature in
SN~2002bo.  This feature was first observed in SN~1994D, where
\citet{meikle1996a} suggested that it comes from either Mg\,{\sc ii}
$\lambda$10926 or He\,{\sc i} $\lambda$10830. However,
\citet{wheeler1998} showed evidence that the line should come entirely
from Mg. \citet{benetti2004a} followed this reasoning and attributed
the feature in SN~2002bo to Mg\,{\sc ii}. The same feature, though
much weaker, is also present in the spectra of SNe~2003du and 2005cf,
but only marginally detectable in SN~1999ee. Whether this variation is a
sign of different Mg abundances or just different ionization and
excitation conditions is unclear. A feature around 16\,800\,\AA\ can 
be seen in all four SNe. \citet{hamuy2002a,hamuy2002b}, 
\citet{marion2003a}, \citet{benetti2004a} and \citet{stanishev2007b} 
agree that it is likely due to Si\,{\sc ii}.  \citet{benetti2004a} 
believe that a weak feature in SN~2002bo at about 20\,800\,\AA\ comes 
from Si\,{\sc ii}. \citet{stanishev2007b} instead propose \SiIII\ as 
the origin of the same feature in SN~2003du. In SN~2005cf the feature 
can also be discerned in the $-10$\,d spectrum, but appears more 
prominent in the $-5$\,d spectrum.

\subsubsection{Spectra around ten days after $B$-band maximum}

Panel (b) of Fig.~\ref{figure:SNe_Ia_NIR_3epochs} shows spectra
between 8.5 and 12.4\,d after $B$-band maximum.

In contrast to the spectra before maximum light, the Mg and Si 
features are no longer present in the spectra 3 weeks later. Instead, 
the NIR spectra now show markedly more structure and are dominated by 
singly ionized Co and Fe. In particular, there are strong Co\,{\sc ii}
features around 15\,500\,\AA\ and 17\,500\,\AA\ in all four SNe.

\subsubsection{Spectra around thirty days after $B$-band maximum}

Panel (c) of Fig.~\ref{figure:SNe_Ia_NIR_3epochs} shows spectra
between 29.4 and 31.5\,d after $B$-band maximum.

The strong features in the $H$ band seem almost identical in the four
Type Ia SNe, but especially similar in SNe~2002bo and 2005cf. In SN~2003du
this region appears to be somewhat better resolved into individual, narrower
features, which could point to a slightly lower maximum expansion
velocity of the iron-group-element-rich core of the ejecta. 
Above 20\,000\,\AA, a characteristic pattern of Co\,{\sc ii} lines 
\citep{wheeler1998} is now very prominent in all four SNe. As shown 
by \citet{marion2009}, these features appear in all SNe~Ia at 
sufficiently advanced epochs. Similarly, a lack of flux in the $J$ 
band, compared to a corresponding black-body continuum, is now 
clearly visible in all the SNe of our sample.

\subsubsection{Mg line velocities}
\label{mg-velocities}

In previous work it was reported that the blueshift of the Mg\,{\sc
  ii} $\lambda$10926 absorption line\footnote{`\MgII\ $\lambda$10926' 
  is a short notation for the triplet\\ 
  $\lambda$10914: 2p$^6$3d $^2$D$_{5/2}$ -- 2p$^6$4p $^2$P$_{3/2}$,\\ 
  $\lambda$10915: 2p$^6$3d $^2$D$_{3/2}$ -- 2p$^6$4p $^2$P$_{3/2}$,\\ 
  $\lambda$10952: 2p$^6$3d $^2$D$_{3/2}$ -- 2p$^6$4p $^2$P$_{1/2}$.} 
does not evolve with time in
SNe~1994D \citep{meikle1996a}, 1999ee \citep{hamuy2002a,hamuy2002b}
and 2003du \citep{stanishev2007b}. \citet{meikle1996a} interpreted
this as the feature being detached, which means that it forms well
above the region of peak emissivity at these wavelengths. Accordingly,
the measured Mg velocity will not change when the emission region
recedes deeper into the core due to the increasing dilution of the
ejecta with time. Thus, the measured \MgII\ velocity traces a zone 
with high \MgII\
abundance, or specifically the lower boundary of such a
zone where the density is highest. 
\MgII\ lines in the optical part 
of the spectrum are less suited to unveil the location of the Mg-rich 
zone since i) they are intrinsically stronger so that very little 
Mg (already on the level of Mg present in the progenitor star) is 
sufficient for them to form \citep{marion2003a}, ii) they are more
strongly blended with lines from other species, and iii) the 
emissivity at shorter wavelengths peaks at much larger radii than 
in the NIR (see Section~\ref{SED}) -- at early phases inside or even 
above the zone of highest \MgII\ abundance.

In order to verify whether the constancy of the \MgII\ $\lambda$10926
velocity may indeed be a generic feature of normal SNe~Ia, we measured
the position of this line in our ISAAC spectra of SN~2005cf. We obtain
velocities of $\sim$\,11\,400 \kms\ for the $-10.2$\,d, $-0.3$\,d and
$+1.8$\,d spectra, and $\sim$\,11\,200 \kms\ for the $+5.4$\,d 
spectrum. Within the typical measurement uncertainties of 
$\sim$\,100--200 \kms\ this is a perfectly constant trend from the
earliest spectrum to the disappearance of the line around maximum
light, supporting the \citet{meikle1996a} interpretation of a detached
feature. The absolute value of the \MgII\ $\lambda$10926 velocity of
SN~2005cf is also quite similar to those derived for SNe~1994D
($\sim$\,11\,700 \kms, \citealt{meikle1996a}) and 1999ee
($\sim$\,10\,500 \kms, \citealt{hamuy2002a,hamuy2002b}), indicating a
similar layering in these three low-velocity-gradient SNe.

Unfortunately there is only a single pre-maximum NIR spectrum available 
for SN~2002bo, and accordingly the time evolution of the \MgII\
$\lambda$10926 velocity cannot be checked. The value of
$\sim$\,14\,400 \kms\ measured in the $-8.5$\,d spectrum, however, is
significantly larger than in the other SNe. SN~2002bo is a high-velocity-gradient SN with
particularly broad and strongly blueshifted \SiII\ and \CaII\ lines in
the early optical spectra. It may well be that the same mechanism that
is responsible for the high-velocity absorptions at optical
wavelengths (a density or abundance enhancement in the very outer
layers, see e.g. \citealt{mazzali2005a}) also acts to produce a NIR
\MgII\ line that is more pronounced and more strongly blueshifted than
usual.

\subsection{UV-optical-infrared}

While UV-optical spectra (with varying quality of the UV part) have
been published for several SNe~Ia
\citep[e.g.][]{kirshner1993a,bufano2009,wang2012a,foley2012a,maguire2012a}
and combined optical-NIR spectra are available for many well-observed
SNe~Ia, complete UV-through-NIR spectra are still a novelty. In fact,
to the authors' best knowledge only a single UV-through-NIR spectrum of
a SN~Ia (though with a very high quality UV part) has thus far seen
the light of day: a maximum-light spectrum of SN~2011iv presented by
\citet{foley2012b}.

\begin{figure}
  \includegraphics[width=84mm]{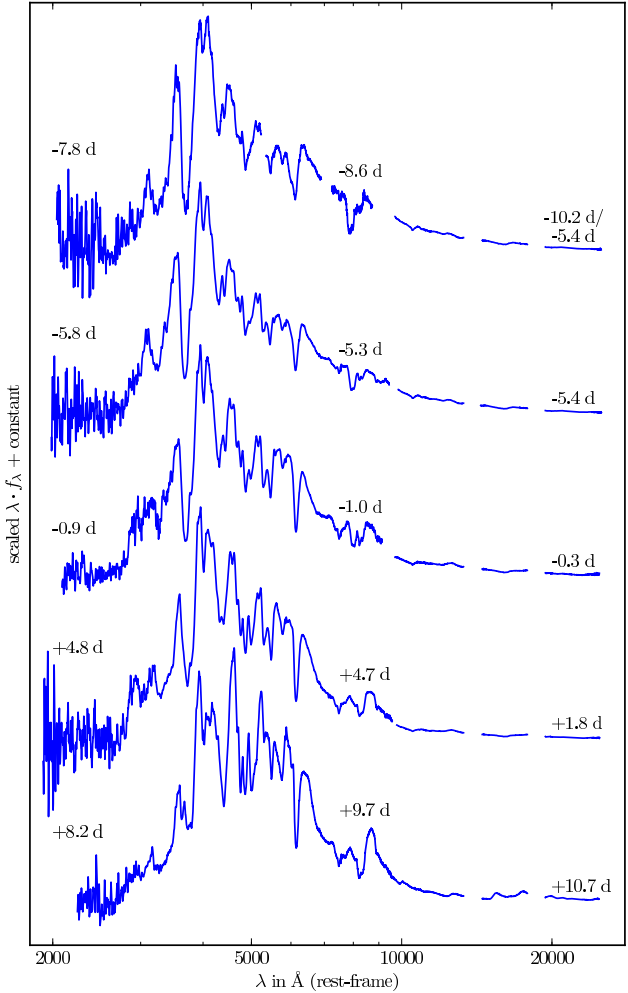}
  \caption{Time sequence of combined UV \citep{bufano2009}, optical 
    \citep{garavini2007} and NIR (this work) spectra of SN~2005cf, 
    scaled to the same peak flux. The NIR spectrum of the first 
    epoch shown in the figure is an average of our $-10.2$\,d and 
    $-5.4$\,d spectra to approximate the spectral appearance at the 
    epoch of the UV-optical spectrum.}
  \label{figure:UVoptIR}
\end{figure}

The unusually complete data set of SN~2005cf in different wavelength
regimes enabled us to create a set of five spectra combined from UV
\citep{bufano2009}, optical \citep{garavini2007} and NIR (this paper)
data (Fig.~\ref{figure:UVoptIR}). As the spectra were typically not
taken at exactly the same epoch, we chose the spectra closest in time
and adjusted the fluxes of the UV and NIR spectra to match the flux
level of the optical spectra. Our UV-through-NIR spectra which cover 
epochs from $\sim$\,8\,d before to $\sim$\,10\,d after maximum light, 
are presented in Fig.~\ref{figure:UVoptIR}.

\citet{foley2012b} already commented on the SED of SN~2011iv at
maximum light being strongly peaked at optical wavelengths. From
Fig.~\ref{figure:UVoptIR} one can see that for SN~2005cf this is true
for all epochs from 8\,d before to 10\,d after maximum, and that the
SED does not change much in this period. The flux blueward of
3000\,\AA\ and redward of 10\,000\,\AA\ never contributes much to the
total emission. The strong UV flux deficit in otherwise blue spectra
is caused by severe line blanketing of iron-group elements at
wavelengths shorter than $\sim$\,4000\,\AA\ \citep{pinto2000b}.  The
main route for UV photons to escape is by fluorescence into the
optical or NIR regime, which is discussed in more detail in
Section~\ref{SED}.

\section{Modelling}
\label{Comparison_SN2005cf_W7}

The discussion of spectral features in the previous section was solely
based on a comparison of our NIR spectra of SN~2005cf to observed
spectra of similar SNe~Ia and line identifications made for those SNe.
Here, we use detailed radiative-transfer models to discuss the
spectral features in the NIR and the evolution of the overall SED in
more detail.

\subsection{Setup}

To this end we used the Monte Carlo radiative-transfer code
\textsc{artis} \citep{kromer2009a,sim2007b} to obtain a
self-consistent solution of the radiative-transfer problem and
calculate synthetic spectra for the hydrodynamical explosion model W7
\citep{nomoto1984a,iwamoto1999a}. W7 is a one-dimensional model of a
carbon deflagration in an accreting Chandrasekhar-mass carbon-oxygen
white dwarf. It is a well-studied and widely-distributed model and
reproduces observed optical spectra and light curves of SNe~Ia
reasonably well (e.g.~\citealt{branch1985a,kasen2006a,
  kromer2009a,jack2011,vanrossum2012a}). In the NIR, however, W7 has
not been studied in great detail yet and the main focus has been laid
on the light curve evolution
\citep{kasen2006a,kromer2009a,jack2012a}. The synthetic NIR spectra of
W7 presented here can thus be a basis for future comparisons among
different radiative-transfer codes.

For our simulation we used the `big\url{_}gf-4' atomic data set of
\citet[][see their table 1]{kromer2009a} with a total of
$\sim8.2\times10^6$ lines. The ejecta were mapped to a $50^3$
Cartesian grid. We then computed the propagation of $10^8$ photon
packets from 2 to 120\,d after the explosion. The calculation was
split into 111 logarithmic time steps. To speed up the initial phase,
a grey approximation, as discussed by \citet{kromer2009a}, was used in
optically thick cells, and the initial 30 time steps (i.e. the initial
six days after the explosion) were treated in local thermodynamic
equilibrium (LTE). Since this is well before the epoch of our first
spectrum of SN~2005cf it should not have any influence on the
following discussion.

\begin{figure}
  \centering
    \includegraphics[width=84mm]{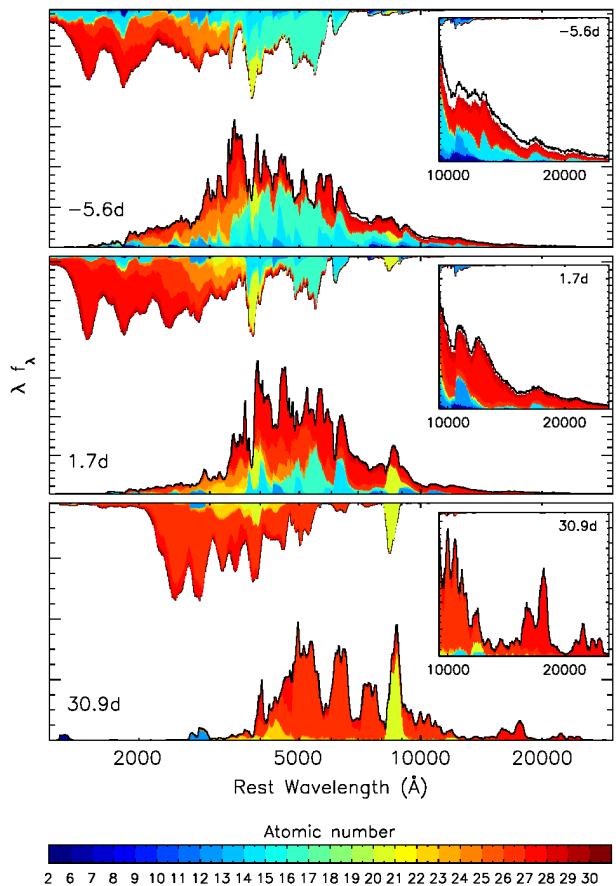}
    \caption{UV to NIR spectral evolution of the W7 model for three
      different snapshots of our radiative-transfer simulation. The
      inlays give an enlarged view to the NIR regime.  Epochs are
      given relative to $B$-band maximum which occurred at 19.5\,d
      after explosion.  The colour coding indicates the elements
      responsible for both bound-bound emission and absorption of
      quanta in the Monte Carlo simulation. The region below the
      synthetic spectrum is colour-coded to indicate the fraction of
      escaping quanta in each wavelength bin which last interacted
      with a particular element (the associated atomic numbers are
      illustrated in the colour bar). Similarly, the coloured regions
      along the top of the plots indicate which elements were last
      responsible for removing quanta from a particular wavelength bin
      (either by absorption or scattering\,/\,fluorescence).
      White regions between the emerging spectrum and the colour-coded 
      bound-bound emission indicate the contribution of continuum 
      processes (bound-free, free-free) to the last emission.}
    \label{figure:fluorescence}
\end{figure}

Three snapshots of the resulting
spectral evolution at $-5.6$, 1.7 and 30.9\,d with respect to $B$-band
maximum (which occurs 19.5\,d after the explosion, at an absolute
magnitude of $-19.14$) are shown in Fig.~\ref{figure:fluorescence} for
UV to NIR wavelengths.

\subsection{Formation of the spectral energy distribution}
\label{SED}

From our Monte Carlo simulation we do not only get the spectral
evolution but also information on the last absorption and emission
processes. In Fig.~\ref{figure:fluorescence} this information is used
to colour-code below the synthetic spectrum the fraction of escaping 
packets in each wavelength bin which were last emitted by bound-bound 
transitions of a particular element (the associated atomic numbers are 
illustrated in the colour bar).  White regions between the emerging 
spectrum and the colour coded part indicate wavelengths at which 
continuum processes (bound-free, free-free) contribute to the last 
emission. The coloured regions along the top of the plots show the 
distribution of photon wavelengths that bound-bound-emitted escaping 
packets had prior to their last interaction. Again the colour coding 
indicates the elements which were responsible for absorbing packets 
in a particular wavelength bin.

Due to energy conservation the total `area' of the emissions and
absorptions, respectively, is equal. Thus the information depicted by
the colour coding in Fig.~\ref{figure:fluorescence} can be used to
understand how the overall SED of our model forms.  Around maximum
light (middle panel) the flux is strongly suppressed by line
blanketing of a dense forest of iron-group-element lines below $\sim$3000\,\AA. 
Between $\sim$3000\,\AA\ and $\sim$6000\,\AA\ prominent absorption
features due to resonance scattering in lines of intermediate-mass
elements are imprinted on an underlying quasi-continuum emitted
by a dense forest of iron-group-element and intermediate-mass-element 
lines. Redward of 6000\,\AA\ the emerging flux is almost only due to 
fluorescent emission -- a prominent exception is the \CaII\ NIR triplet.

As visible from the inlays in the top right corners of the plots in
Fig.~\ref{figure:fluorescence}, this behaviour is even more obvious in
the NIR where almost no absorption occurs.  The only feature which
shows some absorption contribution in this wavelength regime is the
aforementioned feature at 10\,400\,\AA,\ which can be clearly
identified as the \MgII\ $\lambda$10926 line from our simulation. 
Although even this line has a significant contribution from 
fluorescent emission, this confirms the interpretation of 
\citet{wheeler1998} and justifies our use of the associated P-Cygni 
absorption trough in Section~\ref{mg-velocities} to measure ejecta 
velocities of Mg.  Note also that continuum processes contribute to 
some extent to the emission in the NIR (indicated by the white region 
between the emerging spectrum and the colour coded emission due to 
bound-bound processes). From our simulations we find that this 
continuum contribution is dominated by free-free emission.

\begin{figure}
  \centering
  \includegraphics[width=84mm]{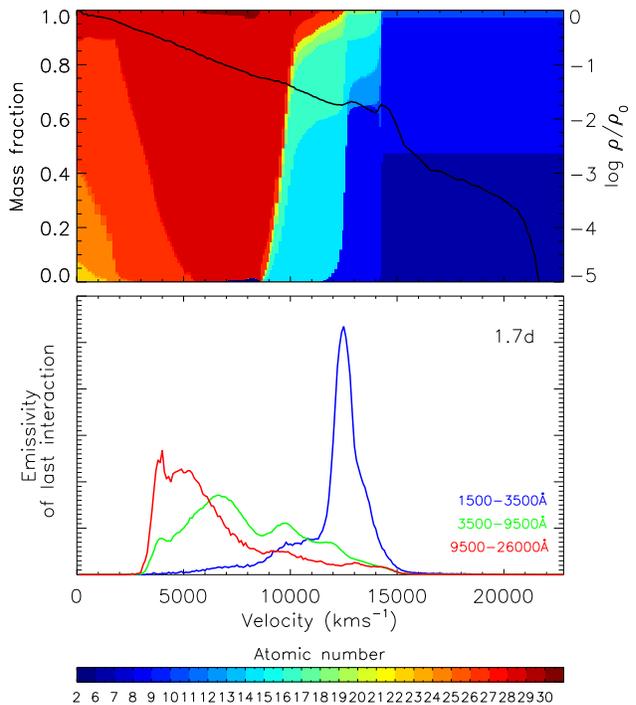}
  \caption{Top panel: abundance structure (colour-coded are the mass 
    fractions of elements, with the associated atomic numbers given in 
    the colour bar at the bottom of the figure) and density profile 
    (black line) of the W7 model. Bottom panel: comparison of the 
    emissivity in the last interaction at UV (1500--3500\,\AA; blue), 
    optical (3500--9500\,\AA; green) and NIR (9500--26\,000\,\AA; red) 
    wavelengths in the W7 model 1.7\,d after maximum light as obtained 
    from our Monte Carlo simulation.}
  \label{figure:emission_depth}
\end{figure}

Comparing the synthetic spectrum at about maximum light (middle panel
of Fig.~\ref{figure:fluorescence}) to pre- and post-maximum epochs
(top and bottom panels, respectively) it is clearly visible that the 
SED is dominated by different species at different epochs.  While
before maximum light intermediate-mass elements dominate the spectral 
features, they are almost negligible at 30.9\,d after maximum (the 
strong NIR triplet of \CaII\ is a prominent exception).  Instead, the 
spectra are then completely dominated by fluorescent emission of 
iron-group elements. This is a consequence of the increasing dilution 
of the ejecta with time, which leads to lower optical depths allowing 
an observer to look deeper into the ejecta (which are strongly layered 
in the case of W7, see top panel of Fig.~\ref{figure:emission_depth}).

In the NIR this transition to an iron-group-element-dominated spectrum 
is already completed at maximum light, and even at pre-maximum epochs 
iron-group-element emission contributes a significant fraction to the 
emerging flux.  This is further illustrated in the histograms in the 
bottom panel of Fig.~\ref{figure:emission_depth}, which show the 
regions of last emission for different wavelength regimes at 1.7\,d 
after maximum light. While the NIR emissivity is strongly peaked at 
$\sim$\,5000\,\kms, the optical emissivity has significant contributions 
from a wider velocity range. The UV, in contrast, originates from a 
narrow region outside the iron-group-element-rich inner core, since the 
dense forest of iron-group-element lines causes extremely large optical 
depths within the core so that UV photons are trapped. The substructure 
in the histograms is a consequence of the emission by different species 
and the strong layering of the ejecta.

\subsection{NIR line identification and comparison to SN~2005cf}

The information on the last absorption and emission processes
presented in the last section cannot only be used to investigate the
formation of the broad-band SED, but also to identify individual features
in the spectra. For that purpose contributions from individual 
ionization states can be extracted from the simulation, similarly 
to the contributions of different elements shown in 
Fig.~\ref{figure:fluorescence}.

\begin{figure*}
  \includegraphics[width=118mm]{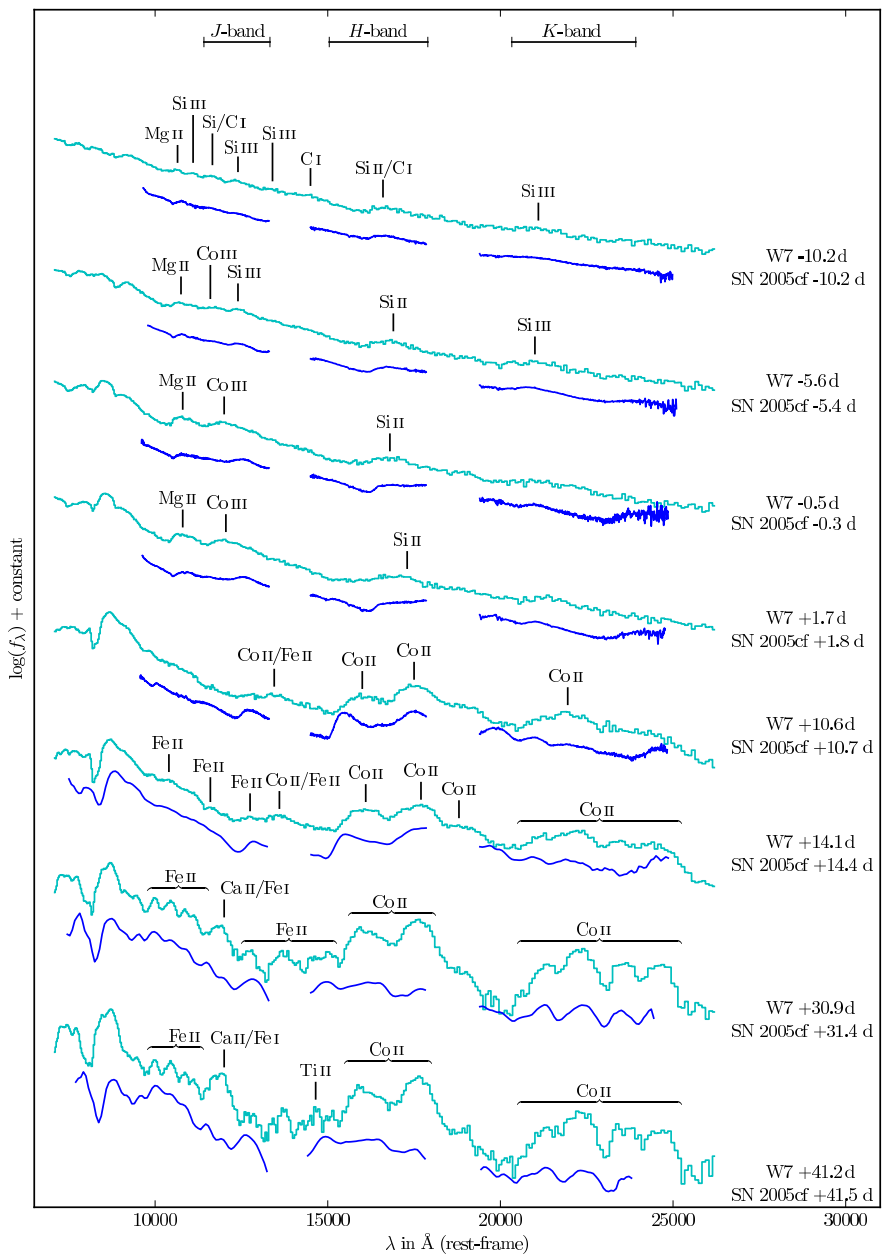}
  \caption{NIR time sequence of our synthetic spectra of the W7 model,
    in comparison with the corresponding NIR spectra of SN 2005cf. An 
    identification of prominent lines in the W7 spectrum as obtained
    from our Monte Carlo simulation is also shown.}
  \label{figure:time_sequence_W7_SN2005cf}
\end{figure*}

We used this approach to identify individual features in the NIR
regime of our model spectra. The results are presented in
Fig.~\ref{figure:time_sequence_W7_SN2005cf} along a time sequence of
our synthetic spectra of the W7 model for 8 epochs between $-10.2$ and
41.2\,d with respect to $B$-band maximum. In order to improve the
signal-to-noise ratio the W7 spectra at 10.2, 5.6, 0.5\,d before and
1.7\,d after peak brightness were obtained by averaging over five,
three, two and two subsequent time steps, respectively. This is not
expected to have a significant influence on the physical information
encoded in the respective spectra, as adjacent time steps are very
close at the earlier epochs due to their logarithmic distribution.

For comparison Fig.~\ref{figure:time_sequence_W7_SN2005cf} also shows
our NIR spectra of SN~2005cf at the corresponding epochs. In general,
the agreement between our synthetic W7 spectra and SN~2005cf is quite
good and it seems that the W7 model, designed only to fit optical
spectra, can also represent NIR spectra at early epochs. That we
identify many more features in the W7 spectrum at $-10.2$\,d than we
can safely detect in SN~2005cf is not necessarily a flaw of W7. It
could rather be due to the fact that these features are relatively
weak (both in the SN and in the simulation) and we are only able to
identify them for the W7 model because we can resolve the various
contributions.

Ten days after $B$-band maximum the agreement between the synthetic
and the observed spectra starts to deteriorate redward of
14\,000\,\AA.\ Although our W7 spectra reproduce the characteristic
\CoII\ features which dominate this wavelength region at about the
right epoch, their relative strengths do not always match those
observed in SN~2005cf. Also there seems to be a small wavelength
offset for the \CoII\ 15\,500\,\AA\ feature. This, for example, might 
be explained by uncertainties in the atomic data. Many of the lines 
used in our simulations which are based on the atomic data of 
\citep{kurucz2006a} are only theoretically predicted so that
both oscillator strengths and resonance wavelengths could be at 
fault,\footnote{For example, for the \CoII\ lines given in 
Table~\ref{rollercoaster} we compared the wavelengths found in the 
atomic data base of \citet{kurucz2006a} with the corresponding 
level energy differences in the NIST database 
(http:/$\!$/www.nist.gov/pml/data/asd.cfm). From 
this we find typical offsets on the order of $\lesssim$\,300\,\kms.}
or some weaker lines may be missing completely. \citet{kasen2006b}
and \citet{kromer2009a} have shown that this can have a significant
influence on NIR observables. Moreover, the simplified excitation 
treatment used by \textsc{artis} (see \citealt{kromer2009a} for a 
detailed description) may lead to inaccurate populations of excited 
levels, particularly at late epochs when non-LTE effects 
become increasingly important.

An interesting result, which will be unaffected by the aforementioned
uncertainties on a qualitative level, is the fact that starting two
weeks after maximum the emission in the various NIR bands is dominated
by different iron-group species. While the $J$-band is dominated by
\FeII\ emission the $H$- and $K$-band emission originates mainly from
\CoII\ (see Figs.~\ref{figure:fluorescence} and
\ref{figure:time_sequence_W7_SN2005cf}). Since the ionisation
potential of both species is of the same order, the $J-H$ and $J-K$
colours are potential probes to the Co/Fe-ratio in SNe~Ia ejecta.
Compared to the observed spectra of SN~2005cf, the W7 spectra have too
much flux in the $H$ and $K$ bands relative to the $J$ band at these
epochs (Fig.~\ref{figure:time_sequence_W7_SN2005cf}). This might
indicate a deficiency of stable iron-group elements in the ejecta of
W7. However, on a quantitative level the uncertainties in the atomic
data and the radiative-transfer treatment currently prevent any firm
conclusions on this issue.

\subsection{Constraining the extension of the iron core}

Like most other features in the NIR spectra of SNe~Ia the four
characteristic \CoII\ humps that dominate the $K$-band spectrum a
month after maximum light are blends. However, our simulation shows
that in this case the number of involved transitions is comparatively
small. In fact, the main contribution comes from no more than 8 \CoII\
emission lines, which are characterised by particularly low-lying
upper levels. Apparently, the population of other \CoII\ levels from
which $K$-band transitions can originate is very low. The wavelengths
and configurations for the 8 \CoII\ lines are given in
Table~\ref{rollercoaster}.

\begin{table}
  \caption{Properties of 8 \CoII\ lines that dominate the $K$-band 
    spectrum of SNe~Ia a month after maximum light, based on the atomic 
    data set of \citet{kurucz2006a}.}
\label{rollercoaster}
\begin{center}
\begin{footnotesize}
\begin{tabular}{ccccccc}
\hline
Wavelength & \multicolumn{3}{c}{Upper level}     & \multicolumn{3}{c}{Lower level} \\
   (\AA)   & Conf.           & Term & $J$  & Conf.        & Term & $J$ \\
\hline
$20\,913$  & 3d$^7$($^4$F)4p & z$^5$F  &  1   & 3d$^6$4s$^2$ & a$^5$D  &  1  \\
$21\,211$  & 3d$^7$($^4$F)4p & z$^5$F  &  2   & 3d$^6$4s$^2$ & a$^5$D  &  2  \\
$21\,351$  & 3d$^7$($^4$F)4p & z$^5$F  &  4   & 3d$^6$4s$^2$ & a$^5$D  &  4  \\
$21\,509$  & 3d$^7$($^4$F)4p & z$^5$F  &  3   & 3d$^6$4s$^2$ & a$^5$D  &  3  \\
$22\,209$  & 3d$^7$($^4$F)4p & z$^5$F  &  5   & 3d$^6$4s$^2$ & a$^5$D  &  4  \\
$22\,482$  & 3d$^7$($^4$F)4p & z$^5$F  &  2   & 3d$^6$4s$^2$ & a$^5$D  &  1  \\
$23\,619$  & 3d$^7$($^4$F)4p & z$^5$F  &  3   & 3d$^6$4s$^2$ & a$^5$D  &  2  \\
$24\,603$  & 3d$^7$($^4$F)4p & z$^5$F  &  4   & 3d$^6$4s$^2$ & a$^5$D  &  3  \\
\hline
\end{tabular} 
\\[1.5ex]
\flushleft
\end{footnotesize}
\end{center} 
\end{table}

We note that the small number of contributing lines makes it possible
to deblend the $K$-band region and derive the extension of the
emitting Co-rich core independently from invoking a complex modelling
machinery. As a simple recipe, this can be accomplished by fitting
Gaussians to each contributing emission line, keeping the central
wavelengths fixed and enforcing a common full width at half maximum
(FWHM) in velocity space. The FWHM which -- besides the individual
intensities -- is a fit parameter, then gives an estimate for the
extension of the emitting Co-rich core. Note, however, that there is
not necessarily a one-to-one correspondence between the two quantities
so these velocities are only indicative. 

Applying this technique to our spectra of SN~2005cf using 8 
Gaussians corresponding to the aforementioned \CoII\ lines 
(Table~\ref{rollercoaster}), we find that the wavelength region
between $\sim$\,20\,800\,\AA\ and $\sim$\,25\,000\,\AA\ is adequately 
reproduced with a FWHM velocity of $\sim$\,11\,000\,\kms\ for the 
spectra at +31 and +41\,d. Slightly smaller FWHM velocities of 
$\sim$\,10\,000 to 10\,200\,\kms\ are derived for the +29\,d spectrum 
of SN~2002bo and the +32 and +42\,d spectra of SN~1999ee.

These numbers are in rough agreement with the extension of the
iron-group-element zone in W7 where those elements are abundant at the
10 per-cent level at a velocity of 10\,600\,\kms, and drop to a mass
fraction of 1 per cent at 12\,300\,\kms. The fair match could already
be expected from the rough agreement between the spectra of SN~2005cf
and the model spectra of W7. It is, however, no validation of the W7
model, since in detail there are numerous other differences between
the model and the observed spectrum.

\section{Conclusions}
\label{Conclusions}

We have presented NIR spectroscopy of SN~2005cf covering epochs
from $-10$\,d to $+42$\,d with respect to $B$-Band maximum. Together
with archival data at other wavelengths this makes SN~2005cf one
of the currently best-observed SNe~Ia and allowed us to compile
combined UV-optical-NIR spectra at five epochs between $\sim$\,8\,d 
before and $\sim$\,10\,d after maximum light. In this respect
SN~2005cf is a unique object since to our best knowledge thus far only 
a single UV-through-NIR spectrum of a SN~Ia has been published by 
\citet{foley2012b} for SN~2011iv.

A comparison of our spectra to observations of SNe~1999ee 
\citep{hamuy2002a,hamuy2002b}, 2002bo \citep{benetti2004a} and 
2003du \citep{stanishev2007b} shows that SN~2005cf is a perfectly 
normal SN~Ia at NIR wavelengths. A similar conclusion has already 
been drawn at optical wavelengths by \citet{garavini2007}, 
\citet{pastorello2007a} and \citet{wang2009}. 

We have also performed radiative-transfer simulations to obtain NIR
spectra for the standard hydrodynamical explosion model W7
\citep{nomoto1984a}. Comparing the obtained synthetic spectra to our
spectra of SN~2005cf we find that W7 reproduces the key properties of
normal SNe~Ia not only at UV-optical as reported in earlier work
\citep[e.g.][]{branch1985a,kasen2006a,kromer2009a,foley2012b} but also
at NIR wavelengths. From a detailed analysis of the emission processes
in the radiative-transfer simulation we have identified the main
spectral features in the NIR. We find only a single clear P-Cygni
line: the \MgII\ $\lambda$10926 triplet which is present at early
epochs. Apart from that the NIR part of the spectrum is almost devoid
of absorption features. Instead, fluorescence is the driving mechanism
that shapes the SED of SNe~Ia.  Before maximum light both
intermediate-mass elements and iron-group elements contribute to the
fluorescent emission. After maximum light almost all the flux
originates from iron-group elements, though different iron-group
species contribute to the various NIR bands. This makes the $J-H$ and
$J-K$ colours a few weeks after maximum light potential probes to the
ratio of different iron-group species. While most of the fluorescent
emission is strongly blended, we find that the characteristic \CoII\
humps which form at about a month after maximum light in the $K$ band
result from a series of relatively unblended lines. This provides a
simple possibility to constrain the iron-group-element-rich regions of
the ejecta from observed spectra.

\section*{Acknowledgments}

The authors thank F. Bufano and V. Stanishev for providing archival 
observations. Additional data were obtained from the Online Supernova
Spectrum Archive (SUSPECT).

This work was supported by the European Union's Human Potential
Programme `The Physics of Type Ia Supernovae' under contract
HPRN-CT-2002-00303, by the Deutsche Forschungs\-gemeinschaft via the
Transregional Collaborative Research Center TRR 33 `The Dark
Universe' and the Excellence Cluster EXC153 `Origin and Structure of
the Universe'. SB is partially supported by the PRIN-INAF 2009 with 
the project `Supernovae Variety and Nucleosynthesis Yields'.

Observations were collected at the ESO 8.2\,m Very Large Telescope UT1
(Cerro Paranal, Chile, programme 075.D-0823) and the Italian 3.58\,m
Telescopio Nazionale Galileo (La Palma, Spain).  We thank the
astronomers at both observatories for their support.  The simulations
were carried out at the John von Neumann Institute for Computing (NIC)
in J\"{u}lich, Germany (project hmu14/hmu20).
Finally, we thank the reviewer for the thorough review and helpful comments.

\bibliographystyle{mn2e}

\label{lastpage}

\end{document}